\documentclass{emulateapj}

\def\ga{\mathrel{\hbox{\rlap{\hbox{\lower4pt\hbox{$\sim$}}}\hbox{$>$}}}}
\def\la{\mathrel{\hbox{\rlap{\hbox{\lower4pt\hbox{$\sim$}}}\hbox{$<$}}}}

\usepackage{verbatim}
\usepackage{amsmath}
\usepackage{multirow}
\usepackage{relsize}
\usepackage{natbib}
\usepackage[flushleft]{threeparttable}
\usepackage{amssymb}
\usepackage{float}

\shorttitle{Anatomy of the PDF}

\shortauthors{Chen, Burkhart, Goodman, \&  Collins}
\begin{document}

\title{The Anatomy of the Column Density Probability Distribution Function (N-PDF)}
\author{Hope Chen\altaffilmark{1}, Blakesley Burkhart\altaffilmark{1}, Alyssa Goodman\altaffilmark{1}, \& David C. Collins\altaffilmark{2}}
\altaffiltext{1}{Harvard-Smithsonian Center for Astrophysics, 60 Garden st. Cambridge, Ma, USA}
\altaffiltext{2}{Department of Physics, Florida State University, Tallahassee, FL 32306-4350, USA}

\begin{abstract}
The column density probability distribution function (N-PDF) of GMCs has been used as a diagnostic of star formation.  Simulations and analytic predictions have suggested the N-PDF is composed of a low density lognormal component and a high density power-law component, tracing turbulence and gravitational collapse, respectively.  In this paper, we study how various properties of the true 2D column density distribution create the shape, or ``anatomy'' of the PDF.  We test our ideas and analytic approaches using both a real, observed, PDF based on Herschel observations of dust emission as well as a simulation that uses the ENZO code.  Using a dendrogram analysis, we examine the three main components of the N-PDF: the lognormal component, the power-law component, and the transition point between these two components.  We find that the power-law component of an N-PDF is the summation of N-PDFs of power-law substructures identified by the dendrogram algorithm.  We also find that the analytic solution to the transition point between lognormal and power-law components proposed by \citet*{Burkhart_2017} is applicable when tested on observations and simulations, within the uncertainties.  We reconfirm and extend the results of \citet*{Lombardi_2015}, which stated that the lognormal component of the N-PDF is difficult to constrain due to the artificial choice of the map area.  Based on the resulting anatomy of the N-PDF, we suggest avoiding analyzing the column density structures of a star forming region based solely on fits to the lognormal component of an N-PDF.  We also suggest applying the N-PDF analysis in combination with the dendrogram algorithm, to obtain a more complete picture of the global and local environments and their effects on the density structures.
\end{abstract}

\section{Introduction}
\label{intro}

Star formation occurs in dense filamentary structures within molecular environments that are governed by the complex interaction of gravity, magnetic fields, and turbulence \citep{McKee_2007}.  The initial distribution of the gas density at parsec scales, which is affected by the average density, level of turbulence and magnetic field strength, may determine the AU scale properties of star formation such as the initial mass function (IMF) and the overall star formation rate \citep{Krumholz_2005,Hennebelle_2011,Padoan_2011b,Padoan_2011a,Federrath_2012,Mocz_2017}.

\subsection{History}
The column density probability distribution function (N-PDF) is a commonly used tool for quantifying the distribution of gas.  Simulations and observations have shown N-PDFs to be an important diagnostic of turbulence and star formation efficiency in local star forming clouds \citep{Federrath_2012, Collins_2010, Burkhart_2015a,Myers_2015}.  This is because N-PDFs can constrain the fraction of dense gas within molecular clouds and provide a means of comparison with analytic models as well as numerical simulations (via synthetic observations) of star formation.

N-PDFs, as available from observations, have been utilized extensively for many different tracers of the ISM.  This includes molecular line tracers such as CO \citep{Lee_2012,Burkhart_2013b} and column density tracers such as dust \citep{Kainulainen_2009,Froebrich_2010,Schneider_2013,Schneider_2014,Schneider_2015b,Lombardi_2015}.  Tracing the N-PDF using dust emission and absorption provides the largest dynamic range of densities, in contrast to molecular line tracers such as CO, which do not trace the true column density distribution due to depletion and opacity effects \citep{Goodman_2009b,Burkhart_2013a,Burkhart_2013b}.

Both the true 3D density (volume density) PDF and N-PDFs have been used to understand the properties of galactic gas dynamics, from the diffuse ionized medium to dense star-forming clouds \citep{Hill_2008,Burkhart_2010,Maier_2017}.  This is because the shape of the density/column density PDF is expected to be related to the underlying physics of the cloud and linked to the kinematics and the chemistry of the gas \citep{Vazquez_1994, Padoan_1997,Kritsuk_2007,Burkhart_2013b,Burkhart_2015a}.

The low column density gas in molecular clouds, as well as in the diffuse neutral and ionized ISM, takes on the form of a lognormal \citep{Vazquez_1994,Hill_2008,Kainulainen_2013}.  This is primarily attributed to the application of the central limit theorem to a hierarchical (e.g. turbulent) density field generated by a multiplicative process, such as shocks.  If the width of the lognormal portion of the N-PDF can be measured, it may be related to the sonic Mach number of the gas in a nearly isothermal cloud \citep{Federrath_2008,Burkhart_2009,Kainulainen_2013,Burkhart_2012,Burkhart_2015a}. 

\begin{figure*}[!b]
\centering
\includegraphics[scale=0.35]{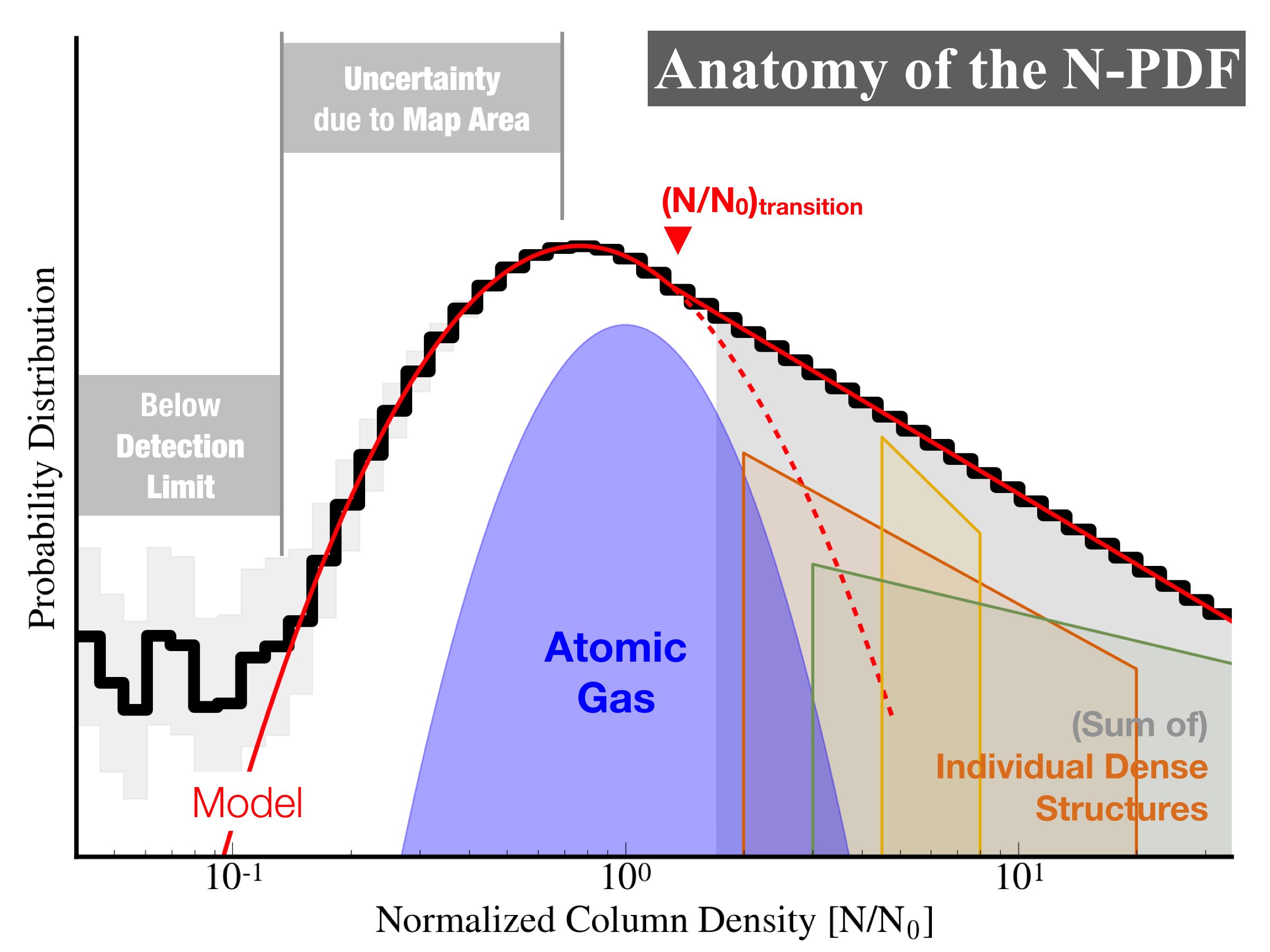}
\caption{The cartoon outlines several components of an N-PDF.  The solid black histogram shows a schematic N-PDF based on observations of a dust tracer, expected of a lognormal + power-law distribution (Equation \ref{eq:piecewise}) and sampled by a histogram.  The red curve shows a perfect lognormal + power-law distribution, denoting a fitted model.  The light gray areas at the lowest column density end indicate the uncertainty in the N-PDF below the sensitivity limit.  In the intermediate column density range, the light gray areas indicate the uncertainty in the N-PDF due to the choice of the map area.  Toward the high column density end, the colored power-law distributions in the power-law regime of the schematic N-PDF are the N-PDFs expected from self-gravitating substructures.  The summation of N-PDFs of all self-gravitating substructures is shown as the gray area in the power-law regime, making up the power-law component of the full N-PDF.  Lastly, the blue-filled lognormal curve denotes the purely lognormal distribution observable with atomic gas tracers such as HI 21cm emission \citep{Burkhart_2015b}.}
\label{fig:cartoon}
\end{figure*}

However, there are known caveats to constraining the exact shape of the lognormal N-PDF.  Below the sensitivity limit, observations are incomplete and do not constitute a statistically meaningful representation of the column density distribution. The sampled N-PDF is then subject to the uncertainty due to the choice of the map area and the foreground/background contamination \citep*[see Figure \ref{fig:cartoon};][]{Lombardi_2015}.  \citet*{Lombardi_2015} found that both changing the map area used to derive the PDF and subtracting the foreground/background uniformly over the map affect the PDF shape at the low column density end (around A$_K$ $\sim$ 0.1, or A$_V$ $\sim$ 1, for nearby molecular clouds).  \citet*{Lombardi_2015} further suggests that the sampling is statistically unbiased only when the region used to sample the PDF is defined by a ``closed contour'' \citep{Alves_2017}.  

In evolved molecular clouds with active star formation, the PDF shape of the densest gas/dust has been observed to develop a power-law form \citep{Kainulainen_2013,Schneider_2015a,Lombardi_2015}.  This is expected from both numerical simulations and the analytic theory which suggest that the PDF of self-gravitating gas should develop a power-law tail \citep{Shu_1977,Federrath_2012,Burkhart_2015a,Meisner_2015}.  The slope of the PDF power law tail depends on self-gravity and the magnetic pressure \citep{Kritsuk_2011,Ballesteros-Paredes_2011,Collins_2012,Federrath_2013,Burkhart_2015a}, and can be analytically related to the powerlaw index of a collapsing isothermal sphere \citep{Shu_1977,Girichidis_2014}.  

Even in the most evolved molecular clouds which exhibit a power law tail towards the dense gas, a lognormal portion of the PDF can still be observed at low total gas and dust column densities.  Recently, the HI PDF in and around GMCs has been measured and shown to be the primary component of the lognormal portion of the total gas+dust PDF \citep{Burkhart_2015b,Imara_2016}.  \citet{Burkhart_2015b,Imara_2016} have shown that the lognormal portion of the column density PDF in a sample of Milky Way GMCs is comprised of mostly atomic HI gas while the power-law tail is built up by the molecular H$_2$, with no contribution from HI.  These studies, including an analytic study by \citet*{Burkhart_2017}, suggest that the transition point in the column density PDF between the lognormal and power-law portions of the column density PDF traces important physical processes.  These include the HI-H$_2$ transition and the so-called ``post-shock density'' regime where the background turbulent pressure equals the thermal pressure and therefore self-gravity becomes dynamically important in the molecular gas \citep*{Burkhart_2017, Li_2015, Kritsuk_2011}.  We illustrate the idealized PDF and the known physics associated with different components in Figure \ref{fig:cartoon}. 
 
\subsection{Understanding Anatomy}
What sets the shape of the N-PDF of a star forming molecular cloud?  
In this paper we seek to address this question by comparing dendrogram-based N-PDFs of observations with simulations that include turbulence, magnetic fields, and self-gravity.

\textit{Dendrograms} are hierarchical tree-diagrams composed of branches, which are split into multiple substructures, and leaves, which have no measurable substructure \citep{Rosolowsky_2008}.  Dendrograms have been applied to star forming turbulent clouds in the past towards understanding the hierarchical properties of star forming regions \citep{Rosolowsky_2008,Goodman_2009a,Beaumont_2012} and supersonic turbulence \citep{Burkhart_2013a}.  We can use dendrograms to break up the column density PDF into its hierarchical constituent parts and relate these individual components to the underlying physics.

In this work, we use the column density PDF derived from dust tracers based on the Herschel observations of the dust thermal emission \citep[the Gould Belt Survey,][]{Andre_2010}, as well as the column density PDF from simulations that include turbulence and self-gravity \citep{Collins_2012,Burkhart_2015a}.  The paper is organized as follows:  
In \S\ref{sec:data} we describe the data we use in this paper, namely Herschel observations of L1689 in Ophiuchus and Enzo MHD simulations, as well as the methods used in our analysis of the N-PDF, including the \textit{dendrogram} algorithm and the fitting to the lognormal and the power-law models.  
We present our analysis of the lognormal component in \S\ref{sec:ln} for the L1689 region in Ophiuchus and for the Enzo simulations.  We then present our dendrogram lognormal + power-law PDF analysis for both simulations and observations in \S\ref{sec:pl}.  We consider the relevance of the analytic solution to the column density at the transition point, as proposed by \citet*{Burkhart_2017}, in \S\ref{sec:obsv_trans} and in \S\ref{sec:sim_trans}, respectively for observations and simulations.  Finally, we discuss our results in \S\ref{sec:discussion} followed by our conclusions in \S\ref{sec:conclusion}.

\section{Data and Methods}
\label{sec:data}
\subsection{Observation}
\label{sec:obsv}
The observed column density for the L1689 region in Ophiuchus (Oph L1689) is derived from data taken by the \textit{Herschel Space Observatory}.  Herschel was a satellite operated by the European Space Agency.  Its Photodetecting Array Camera and Spectrometer (PACS) and Spectral and Photometric Imaging Receiver (SPIRE) covered wavelengths from 55 to 670 $\mu$m, with six broad spectral bands identified by their nominal central wavelengths: PACS 70 $\mu$m, 100 $\mu$m, and 160 $\mu$m, and SPIRE 250 $\mu$m, 350 $\mu$m, and 500 $\mu$m.  The data used to derive the column density presented in this paper were obtained as part of the Herschel Gould Belt Survey \citep{Andre_2010}.

To derive the column density, we make use of the maps at 160 $\mu$m, 250 $\mu$m, 350 $\mu$m, and 500 $\mu$m, produced by the Herschel Interactive Processing Environment (HIPE; Version 11.1.0).  The maps produced by HIPE were not absolutely calibrated, in the sense that there may be a per-band additive offset needed to correct the Herschel zero level.  This issue has traditionally been addressed for small $\lesssim$1 square degree regions by adding a scalar offset to the Herschel mosaic under consideration at each wavelength.  However, in this study we seek to calibrate Herschel data for a large region of the Ophiuchus cloud dozens of square degrees in size. Over this sizeable footprint, we found that a single per-band scalar offset could not satisfactorily correct the Herschel zero level.

Instead, we chose to allow for a spatially varying zero-level offset in each Herschel band.  Specifically, we used the \cite{Meisner_2015} \emph{Planck}-based thermal dust emission model to predict the Herschel emission at 10$'$ FWHM over our entire Ophiuchus footprint.  These low-resolution predictions incorporated color corrections to account for the Herschel bandpasses.  To correct the Herschel zero level, we then high-pass filtered each Herschel mosaic at $10$$'$, and replaced the low-order spatial modes ($\geq$10$'$ FWHM) with the corresponding Planck-based predictions.  We thereby achieved dust emission maps at 160$\mu$m, 250$\mu$m, 350$\mu$m, 500$\mu$m which retain the high angular resolution of Herschel, but inherit the reliable zero level of Planck.

To derive reliable column density maps, we assume that the Herschel maps from 160$\mu$m to 500$\mu$m, following the zero level corrections, are dominated by ``big grain \citep[BG;][]{Stepnik_2003}'' thermal dust emission.  We also assume that the thermal dust emission can be described by a single modified blackbody (MBB).  This assumption is only valid under certain conditions and becomes inaccurate at wavelengths shorter than 100$\mu$m due to contamination from ``very small grain'' (VSG) emission.  Adopting a single-component modified blackbody model also presumes that the material along the line of sight can be characterized well by a single temperature.  We incorporate a spatially varying value of dust emissivity power-law index, $\beta$ in the following SED fitting.

We smooth the Herschel maps at 160$\mu$m, 250$\mu$m and 350$\mu$m to 36.1$'$ FWHM, to match the angular resolution of the SPIRE 500$\mu$m map.  In the SED fitting, we assume a 10\% fractional uncertainty on each Herschel intensity measurement.  For each pixel, we then derive the optimal temperature and 350$\mu$m intensity via simple $\chi^2$ minimization.   Using the equation $I_{\nu}$=$\tau_{\nu}$$B_{\nu}(T)$, we can then derive the optical depth at any frequency based on our two fitted parameters.  Then, a conversion from the 350$\mu$m optical depth to column density units are obtained by convolving the map of the 350$\mu$m optical depth ($\gamma_{350}$) to the nominal resolution of the NICEST map \citep{Lombardi_2009} and comparing to the NICEST extinction map (in the unit of K-band extinction magnitude, $A_K$).  A simple power law, $A_K$ = $\gamma\tau_{350}^{c}$, is fitted for data points around the median value of 350$\mu$m optical depth, $\sim$ 2.5$\times10^{-4}$.  The resulting parameters, $\gamma$ $\sim$ 2520 and $c$ $\sim$ 1.11, are consistent with the solution suggested by \citet[$\gamma$ = 2500 with an \emph{almost perfect linearity}, see also Eq.11-14;][]{Lombardi_2014}.

\begin{figure*}[!b]
\centering
\includegraphics[scale=0.45]{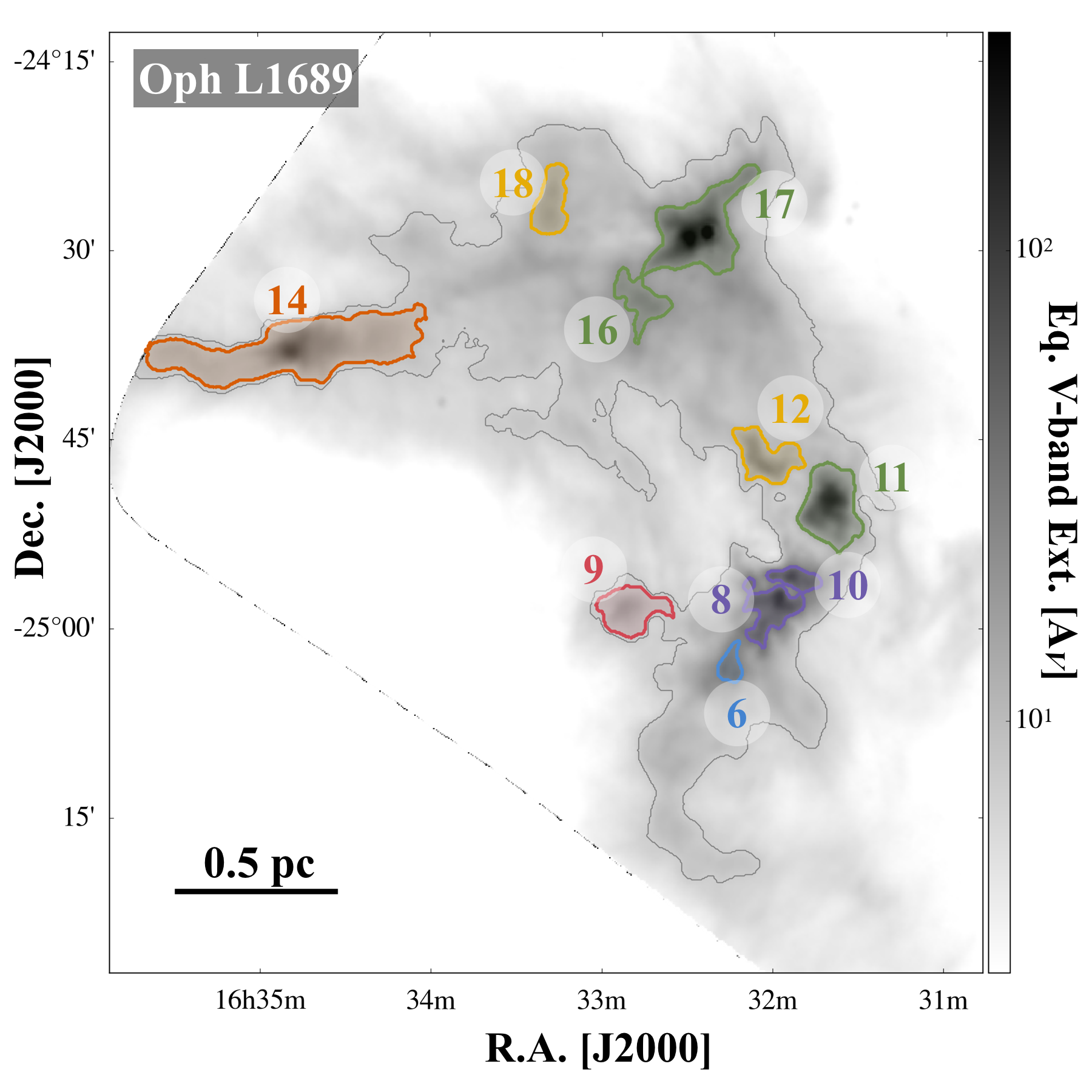}
\caption{\label{fig:map_obsv}
The column density map of L1689 in Ophiuchus.  The colored regions show the leaf structures in the dendrogram.  (The N-PDFs of the leaf structures are in Figure \ref{fig:fancy_obsv}.)  The gray contour corresponds to N/N$_0$ = 0.95, which is used as the minimum value (\texttt{min\char`_value} in \texttt{astrodendro}; see \S\ref{sec:dendro}) in the dendrogram analysis.  N$_0$ is $\sim$ 7.3 mag in the unit of equivalent V-band extinction ($A_V$) for L1689.}
\end{figure*}

Figure \ref{fig:map_obsv} is the final column density map of the L1689 region.  The map covers a region of $\sim$ 2.5 pc by 2.5 pc, and includes all material with column density larger than $A_K$ = 0.8 mag \citep[used to define the ``dense molecular cloud'' where the star formation almost exclusively occurs][]{Lada_1992,Lada_2010}.

The resulting column density measurements can then be converted to the unit of equivalent V-band extinction magnitudes, using $A_V$ = $A_K$/0.112 \citep{Rieke_1985}, and to the number column density, using N(H$_2$)/$A_V$ = 9.4$\times$10$^{20}$ cm$^{-2}$ \citep{Bohlin_1978} or N(H$_2$)/$A_V$ = 6.9$\times$10$^{20}$ cm$^{-2}$ \citep{Draine_2003,Evans_2009}.  In this paper, for easier comparison with other works in various column density units and between simulation and observation, the column density is expressed in a dimensionless, normalized unit where the column density is divided by the median value above the detection limit.  For L1689, the median column density is $\sim$ 7.3 mag in the unit of equivalent V-band extinction ($A_V$).  Figure \ref{fig:map_obsv} shows the map used in the following analysis, and Figure \ref{fig:cartoon_obsv} shows the N-PDF of the entire L1689 region in the normalized units.  Note that the conversion from physical to dimensionless units does not change the shape of the N-PDF on the logarithmic scale.

An overview of physical properties of Oph L1689 is listed in Table \ref{table:comparison}, in comparison to the Enzo simulation.

\subsubsection{Turbulence in Oph L1689}
\label{sec:obsv_turb}
To estimate the magnitude of turbulent motions in Oph L1689, we fit Gaussian profiles for spectra from the FCRAO observations of $^{13}$CO (1-0) molecular line emission \citep[using data from the COMPLETE Survey,][]{Ridge_2006}.  Assuming that the dust temperature derived from Herschel observations of thermal dust emission is representative of the gas temperature (\textit{i.e.} the gas and the dust are in thermal equilibrium; see \S\ref{sec:obsv} for details on the fitting of dust properties), we calculate the sonic Mach number for Oph L1689, $\mathcal{M}_s$ = 4.50$^{+1.43}_{-1.09}$.  Since the $^{13}$CO (1-0) line emission traces a density range similar to the density range (as traced by thermal dust emission) in question in this paper, we use the Gaussian line widths of the $^{13}$CO (1-0) transition and the estimated sonic Mach number to assess the dynamics of Oph L1689 in the following analyses (see \S\ref{sec:obsv_pl}).  The estimated sonic Mach number is also used to calculate the transitional column density (\S\ref{sec:obsv_trans}), between the lognormal and the power-law components, in the analytic model proposed by \citet*{Burkhart_2017}.  See \S\ref{sec:comparison} and Table \ref{table:comparison} for a comparison of physical properties to the Enzo simulation used in this paper.

\subsection{Simulation}
\label{sec:sim}
In order to investigate the physics of the N-PDF without the uncertainties that plague observations, we use MHD simulations of a collapsing turbulent cloud produced by the Enzo code \citep{Collins_2010,Collins_2012}.  The Enzo simulations used in this paper are generated by solving the ideal MHD equations with large-scale solenoidal forcing \citep[$b$ $\approx$ 1/3, where $b$ is the forcing parameters; see \S\ref{sec:trans} in this paper, and][]{Collins_2012,Burkhart_2017}.  The simulations have a sonic Mach number of $\mathcal{M}_s$ = 9 and an Alfv\'enic Mach number of $\mathcal{M}_A$ = 12, which scales to $\sim$ 4.4 $\mu$G assuming a sound speed of $c_s$ = 0.2 km s$^{-1}$ and an average volume density of $n_H$ = 1000 cm$^{-3}$ \citep*[the ``mid'' case in][]{Burkhart_2015a}.  To investigate properties of the column density profile under the influence of gravitational collapse, the snapshot at the simulation time t = 0.6 t$_\text{free-fall}$ is taken.  Each side of the simulation cube is $\sim$ 4.6 pc in length, and the density and the velocities are sampled on a 512$^3$ grid (resulting in a coarser grid cell size of $\sim$ 9$\times$10$^{-3}$ pc, or $\sim$ 1.8$\times$10$^3$ AU; compared to the smallest physical scale the simulation resolves with adaptive mesh refinement of $\sim$ 500 AU).

To obtain the column density, the density cube is integrated along one of the three axes of a density cube, through the full 4.6 pc.  The resulting column density map is then convolved with a Gaussian beam with the same size as that of the Herschel 500-$\mu$m beam.  For subsequent analyses involving the \textit{dendrogram} algorithm, we take a 2D map region of 2.3pc by 2.3pc, in order to compare to Oph L1689.  Similarly, the column density is expressed in the normalized units, where the column density is divided by the median value.  (See \S\ref{sec:obsv} for details about the ``normalization''.)  The median value of the simulated column density map is $\sim$ 1.3$\times$10$^{22}$ cm$^{-2}$, or $\sim$ 13.5 mag in the unit of equivalent V-band extinction ($A_V$).  See \citet{Collins_2012} and \citet*{Burkhart_2015a} for details on scalings to physical units.

An overview of physical properties of the Enzo simulation is listed in Table \ref{table:comparison}, in comparison to Oph L1689.

\subsubsection{Turbulence in the Enzo simulation}
\label{sec:sim_turb}
\begin{figure*}[!b]
\centering
\includegraphics[scale=0.45]{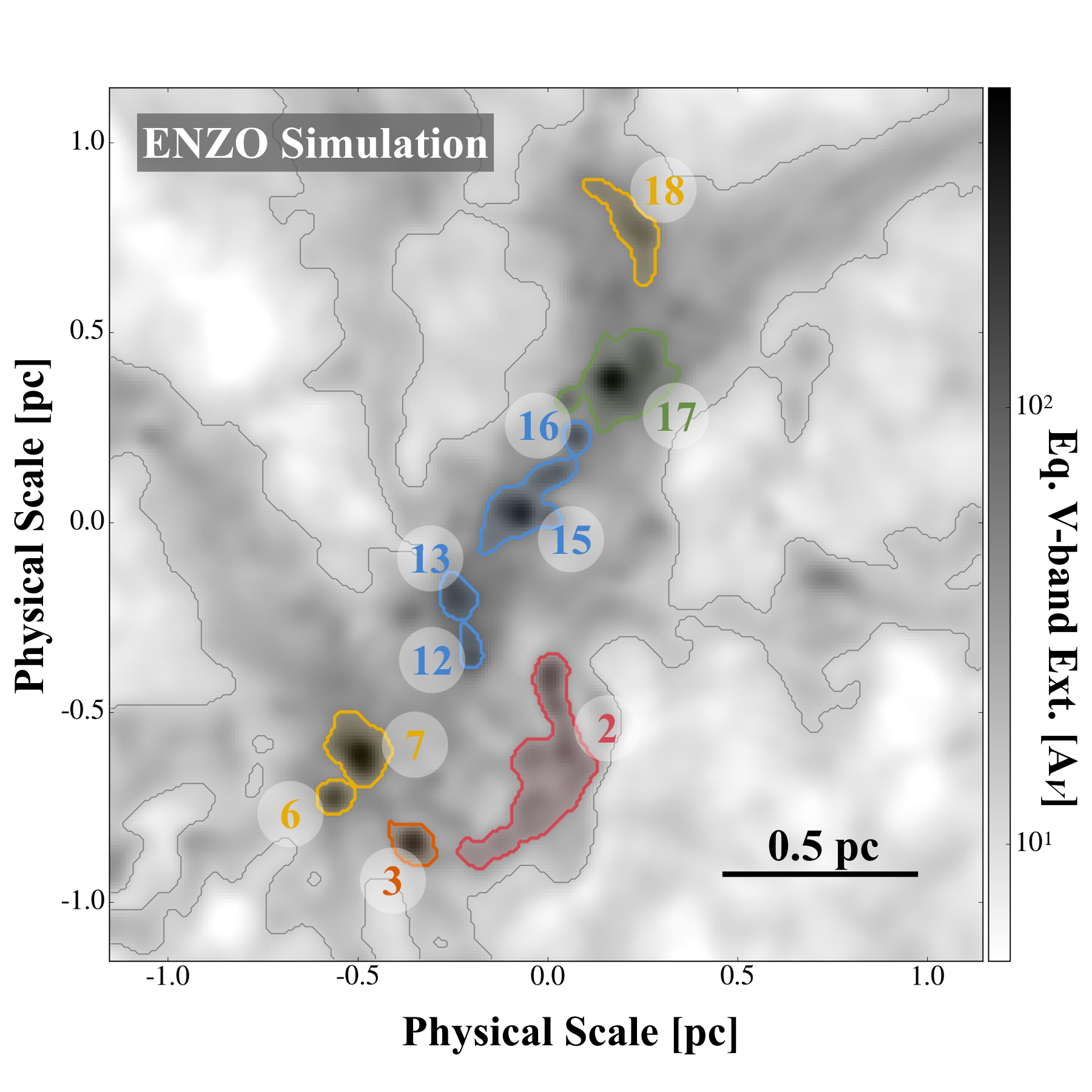}
\caption{\label{fig:map_sim}
The map of the integrated number density at t=0.6t$_{ff}$ in the Enzo simulation.  The colored regions show the leaf structures in the dendrogram.  (The N-PDFs of the leaf structures are in Figure \ref{fig:fancy_sim}.)  The gray contour corresponds to N/N$_0$ = 0.95, which is used as the minimum value (\texttt{min\char`_value} in \texttt{astrodendro}; see \S\ref{sec:dendro}) in the dendrogram analysis.  N$_0$ is $\sim$ 13 mag in the unit of equivalent V-band extinction ($A_V$) assuming scalings in \citet{Collins_2012}.}
\end{figure*}

To estimate the magnitude of turbulence in the 2.3pc by 2.3pc by 4.6pc cube (a 2.3pc by 2.3pc map with a 4.6pc line of sight) from which the column density map (Figure \ref{fig:map_sim}) was derived, we calculate the sonic Mach number, $\mathcal{M}_s$, based on the 3D velocity dispersion in the 2.3pc by 2.3pc by 4.6pc cube.  Since we chose a region where gravity is particularly dominant (see Figure \ref{fig:map_sim}, in which multiple high-density structures are identifiable), we expect a smaller sonic Mach number (less turbulent material) than expected of the entire 4.6pc by 4.6pc by 4.6pc cube \citep[$\mathcal{M}_s$ = 9;][]{Collins_2012}.  For the 2.3pc by 2.3pc column density map, we find $\mathcal{M}_s$ = 8.14$^{+2.95}_{-1.81}$.  The estimated sonic Mach number is used to calculate the transitional column density in the analytic model proposed by \citet*{Burkhart_2017}.  See \S\ref{sec:trans} for the analysis of the analytic model.

\subsubsection{Difference between observation and simulation}
\label{sec:comparison}
Table \ref{table:comparison} shows that Oph L1689 and the Enzo simulation used in this paper have different median column densities and sonic Mach numbers, indicating that Oph L1689 and the Enzo simulation are at different stages of star formation and/or under the effects of different levels of gravity, turbulence, and likely also magnetic field.  But, since the main goal of this paper is to examine the uncertainties in the N-PDF analysis and the origins of the different components of the N-PDF, we are more concerned with the relative column density structures than the absolute dynamics of Oph L1689 and the Enzo simulation.  The dendrogram analyses of Oph L1689 and the Enzo simulation show that the two \textit{do} have similar hierarchical density structures (see Table \ref{table:comparison}, and compare Figure \ref{fig:fancy_obsv} to Figure \ref{fig:fancy_sim}).  As a side note, \citet{Beaumont_2013} have also demonstrated that it remains difficult to make simulations ``look like'' a real molecular cloud, even in terms of the simplest observable diagnostics including the column density distribution and the distribution of the CO line widths.

\begin{table*}[ht]
\caption{\label{table:comparison}Comparison between Oph L1689 and the Enzo simulation.}
\centering
\begin{tabular}{lcccc}
 & Median Column Density  & Sonic Mach Number & \multicolumn{2}{c}{Dendrogram properties\tablenotemark{a}} \\
 \cline{4-5}
 & N$_0$ [$A_V$] & $\mathcal{M}_s$ & Number of Levels & Number of Leaf Structures \\
\hline
\hline
Observation (Oph L1689) & 7.3$\pm$3.5 & \tablenotemark{b}4.50$^{+1.43}_{-1.09}$ & 8 & 10 \\
Simulation (Enzo) & 13.5$\pm$4.7 & \tablenotemark{c}8.14$^{+2.95}_{-1.81}$ & 8 & 10 \\
\end{tabular}
\begin{tablenotes}
\item[a]{\textbf{a.} The dendrograms are computed in normalized units, N/N$_0$.}
\item[b]{\textbf{b.} The Mach number is derived from the average Gaussian line widths for the $^{13}$CO (1-0) molecular line emissions in Oph L1689.}
\item[c]{\textbf{c.} The Mach number is calculated for the 2.3pc by 2.3pc by 4.6pc cube, from which the 2.3pc by 2.3pc column density map shown in Figure \ref{fig:map_sim} is derived.}
\end{tablenotes}
\end{table*}

\subsection{Dendrogram}
\label{sec:dendro}
We use \textit{dendrograms} \citep{Rosolowsky_2008} to identify substructures inside molecular clouds.  The \textit{dendrogram} is a structure finding algorithm that decomposes an N-dimensional (N $\geq$ 1) entity into smaller \textit{substructures} based on the positions and the values of the pixels.  The \textit{dendrogram} algorithm can build a tree of substructures representative of the hierarchy of nested column density features inside a cloud \citep[for previous examples of application in astrophysics and a detailed description of the algorithm, see][]{Rosolowsky_2008,Goodman_2009a,Burkhart_2013a}.  This hierarchical representation is ideal for the following analysis, where we hope to find the composition of various components of the N-PDF.

\begin{figure*}[!b]
\centering
\includegraphics[scale=0.35]{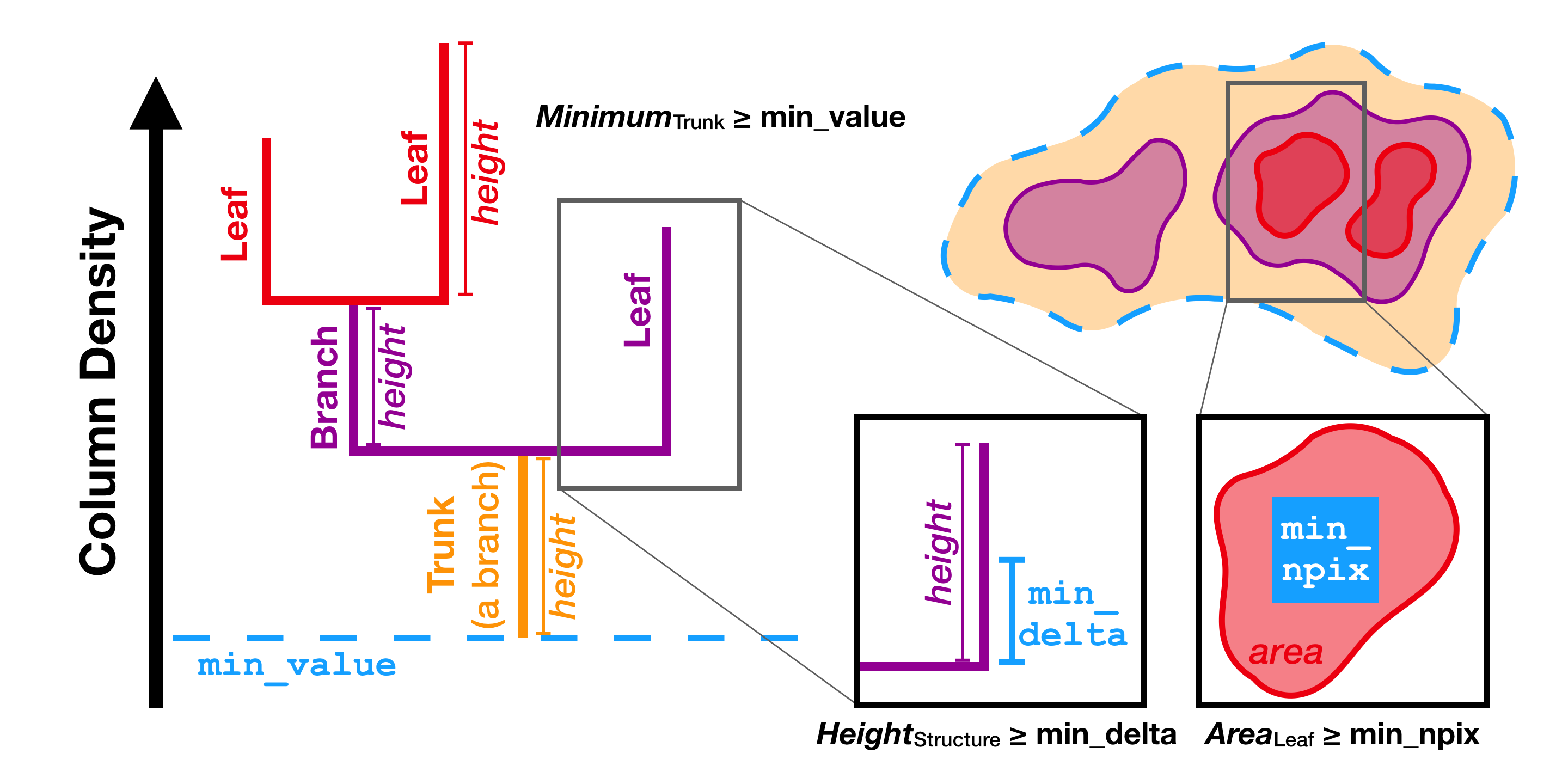}
\caption{This cartoon shows how the three criteria used in the computation of a dendrogram, which are described in \S\ref{sec:dendro}, are defined.  On the left-hand side of the cartoon, a dendrogram is shown using a vertical axis of column density.  The \textit{branch}, \textit{leaf}, and \textit{trunk} structures and their heights are labeled according to their definitions (see \S\ref{sec:dendro}).  On the top right is a cartoon 2D map from which the mock dendrogram on the left-hand side is derived.  Contours on the 2D map define the boundaries of dendrogram substructures of corresponding colors.  The three criteria---1) the minimum value of a trunk structure has to be larger than \texttt{min\char`_value}, 2) the height of any structure in the dendrogram has to be larger than \texttt{min\char`_delta}, and 3) the area of a leaf structure has to be larger than \texttt{min\char`_npix}---are shown in light blue with bold-faced annotations.  See \S\ref{sec:dendro} for details on the definitions.}
\label{fig:dendrogram}
\end{figure*}

Figure \ref{fig:dendrogram} shows a cartoon demonstrating how a dendrogram is computed from a two-dimensional map.  The \textit{dendrogram} (shown on the left-hand side of Figure \ref{fig:dendrogram}) is a tree diagram where each structure can harbor exactly two substructures.  The structures can then be categorized into \textit{leaf}, \textit{branch}, or \textit{trunk} structures.  The \textit{trunk} structure is a special case, as it is the bottommost structure in the dendrogram.

We define the \textit{height} of any structure (leaf, branch, or trunk) to be its vertical extent as shown in Figure \ref{fig:dendrogram}.  Three input parameters effect the outcome of the dendrogram algorithm.  First, the minimum value in a \textit{trunk} structure has to be larger than \texttt{min\char`_value}\footnote{\label{fn:astrodendro}\texttt{min\char`_value}, \texttt{min\char`_delta}, and \texttt{min\char`_npix} are names of the input parameters in the \textit{Python}-based \texttt{astrodendro} package.  In this paper, we use them as shorthands for the three input parameters in the dendrogram algorithm.  See \url{http://dendrograms.org} for documentation of \texttt{astrodendro}.} (the light blue dashed line in the dendrogram on the left and the light blue dashed contour in the 2D map on the top right of Figure \ref{fig:dendrogram}).  Second, the height of any substructure in the dendrogram has to be larger than \texttt{min\char`_delta} (the left inset diagram on the bottom right of Figure \ref{fig:dendrogram}).  Lastly, the area of a \textit{leaf} structure must be larger than \texttt{min\char`_npix} (the right inset diagram on the bottom right of Figure \ref{fig:dendrogram}).

In this paper, we use the \textit{Python}-based \texttt{astrodendro} package\footnotemark[3] to compute and analyze the dendrograms of the column density maps.  (See \S\ref{sec:obsv} and \S\ref{sec:sim} for details on how we derive the column density maps from observations and simulations.)  The \texttt{astrodendro} package offers a simple control of the three essential parameters---\texttt{min\char`_value}, \texttt{min\char`_delta}, and \texttt{min\char`_npix}---in the dendrogram analysis, each defining one of the three criteria described in the above paragraph (see Figure \ref{fig:dendrogram}).  We use the same set of parameters in normalized units for observations and simulations in order to make sure that the analyses are consistent.  We choose a minimum value (\texttt{min\char`_value}) of N/N$_0$ = 0.95, which corresponds to $A_K$ $\sim$ 0.8 mag in observations of Oph L1689.  Note that $A_K$ = 0.8 mag is also found to define the ``dense cloud'' where the star formation activity occurs \citep*{Lada_2010}.  We then choose \texttt{min\char`_npix} that corresponds to an area equivalent to a 0.05pc by 0.05pc square.  And to guarantee a representative sampling of the N-PDF, we also make sure that each substructure has more than 200 pixels \citep*[for discussions on the sampling, see][]{Clauset_2009}.  Lastly, \texttt{min\char`_delta} is selected to be N/N$_0$ = 0.475, roughly corresponding to the 3-$\sigma$ level in observations.  The final dendrograms are shown as connections between panels in Figure \ref{fig:fancy_obsv} and Figure \ref{fig:fancy_sim}.

\subsection{Fitting the N-PDF}
\label{sec:fitting}
In this paper, the term ``N-PDF'' is used to indicate the frequency distribution of the column density in a map or a binned histogram representation of this frequency distribution.  This definition is mathematically different from a well-defined probability distribution, where the distribution function $f(x)$ is normalized ($\int_0^\infty{f(x)dx}$ = 1).  However, this discrepancy in definition does not affect the validity of results presented in this paper.  In practice, since one can only sample the mathematical probability distribution function with a finite number of independent pixels (in both observations and simulations), the results in this paper can be directly compared to other work where the distributions of values in column density maps are examined.

Following results presented by \citet{Schneider_2013,Myers_2015} and \citet*{Burkhart_2017} and the anatomical diagram shown in Figure \ref{fig:cartoon}, we assume that the N-PDF has a lognormal component in the low column density regime and a power-law component toward the high column density end.  The two components are continuous at the transitional column density \citep*{Myers_2015,Burkhart_2017}.  By defining the normalized column density on the logarithmic scale:

\begin{equation}
s \equiv \ln{\text{N}/\text{N}_0},
\end{equation}

\noindent the distribution can be written as

\begin{equation}
\label{eq:piecewise}
p_s(s) =
\begin{cases}
	M\frac{1}{\sqrt{2\pi}\sigma_s}\exp{\left[-\frac{\left(s-s_0\right)^2}{2\sigma_s^2}\right]},& s < s_t\\
    Mp_0\exp{\left[-\alpha s\right]},& s > s_t,
\end{cases}
\end{equation}

\noindent where $s_t$ is the transitional column density in the logarithmic normalized units, $p_0$ is the amplitude of the N-PDF at the transition point, and $M$ is the normalization/scaling parameter.  To carry out the least-squares fitting, the N-PDF is first sampled using a binned histogram.  The $\chi^2$-residual between the model and the histogram is minimized over a range of $s_t$, with a step size in $s_t$ equal to one third of the histogram bin size.  For convenience, the modeled N-PDF based on this equation is called a ``lognormal + power-law'' model in discussions below.

We note that the least-squares fitting cannot be used to determine whether a lognormal + power-law model is a better fit to the observed distribution than a simple lognormal model \citep*{Clauset_2009} and is subject to various fitting/binning choices.  Thus, we limit our analyses involving $\chi^2$-fitting to the N-PDFs of the \emph{cloud-scale} regions (the entire L1689 region or the integrated column density maps derived from the simulation cubes; see \S\ref{sec:obsv} and \S\ref{sec:sim} for how we define the regions in observations and simulations).  At the cloud scale, the large number of independently sampled pixels (on the order of 10$^4$ to 10$^5$) makes the $\chi^2$-fitting results less dependent on fitting/binning choices \citep{Stutz_2015}, and the existence of a power-law component is evident by eye (see Figure \ref{fig:cartoon_obsv} and Figure \ref{fig:cartoon_sim}).  For smaller substructures within the cloud, we base our analysis on comparing (without fitting) the individual N-PDFs to the N-PDFs of entire regions.  Thus, the results presented in this paper are essentially independent of the uncertainties in fitting.  (See \citet{Stutz_2015} for a comparison of various fitting/binning schemes, and \citet*{Burkhart_2015a}, for alternative N-PDF diagnostics that do not require fitting.)

\section{The Lognormal Component}
\label{sec:ln}

\subsection{Observation}
\label{sec:obsv_ln}
\begin{figure*}[!b]
\centering
\includegraphics[scale=0.35]{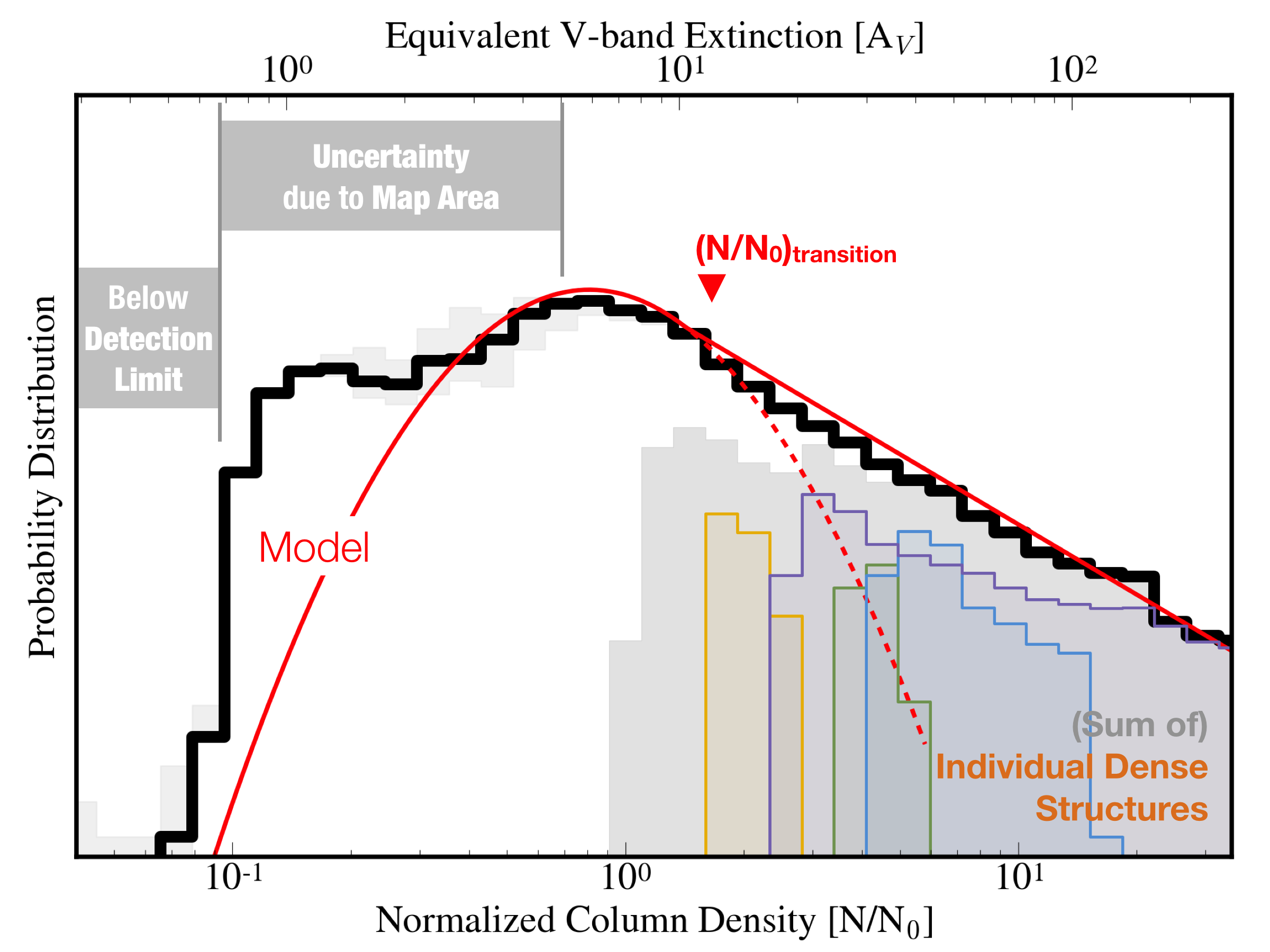}
\caption{\label{fig:cartoon_obsv}
Similar to Figure \ref{fig:cartoon}, but using the real Herschel-based column density map of Oph L1689 instead.  The solid black histogram represents the N-PDF of the Hershcle-based map.  The red curve is a lognormal + power-law fit to the histogram above the range of column density affected by the uncertainty due to the choice of the map area and the detection limit.  The light gray area toward the low column density end shows the change in N-PDF following the change in the map area by $\pm$20\%.  Toward the high column density end, the colored power-law distributions in the power-law regime of the full N-PDF are the N-PDFs of independent ``leaf'' structures in the dendrogram.  The summation of N-PDFs of all ``leaf'' substructures is shown as the gray area in the power-law regime, making up most of the power-law component of the full N-PDF.}
\end{figure*}

We investigate the lognormal component of the N-PDF of Oph L1689 and plot the total PDF in Figure \ref{fig:cartoon_obsv}.  Figure \ref{fig:cartoon_obsv} shows that the lognormal component is observable only between N/N$_0$ $\sim$ 0.6 and $\sim$ 1.4 (the transition point; $A_V$ $\sim$ 4.4 mag and $\sim$ 10.2 mag).  Figure \ref{fig:cartoon_obsv} also demonstrates that sampling of the lognormal component of the N-PDF is subject to the uncertainty due to the map area between N/N$_0$ $\sim$ 0.1 and $\sim$ 0.6 \citep*[$A_V$ $\sim$ 0.7 mag and $\sim$ 4.4 mag, consistent with][]{Lombardi_2015}, and that the lognormal component is unobservable below the detection limit (N/N$_0$ $\sim$ 0.1 in this case).  The uncertainty due to the map area and the detection limit leaves us with a very narrow range of well-sampled column density that can be used in an attempt to fit for the lognormal component.  In the case of Oph L1689, the range of column density that is not affected by the uncertainties is $\lesssim$ 50\% of the full range the fitted lognormal component spans \citep[which includes the pixels outside the last closed contour;][]{Alves_2017}.  Thus, a direct comparison of the width of the lognormal component to the cloud dynamics such as the sonic Mach number is \emph{difficult and unreliable} using the column density maps based on dust tracers.

\subsection{Simulation}
\label{sec:sim_ln}
\begin{figure*}[!b]
\centering
\includegraphics[scale=0.35]{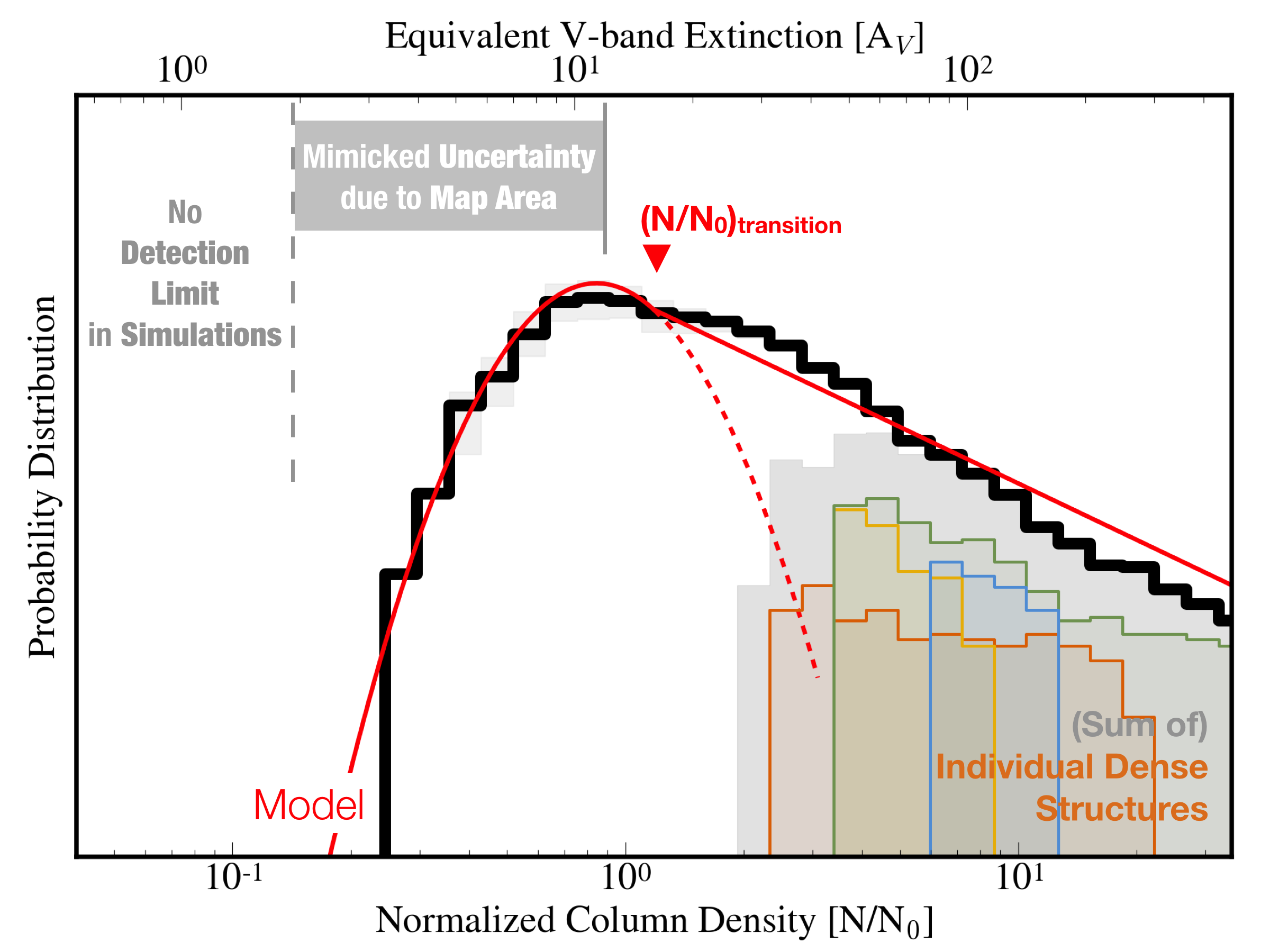}
\caption{\label{fig:cartoon_sim}Similar to Figure \ref{fig:cartoon}, but using the integrated column density map based on the Enzo simulation (see \S\ref{sec:sim}).  The solid black histogram represents the N-PDF of the entire column density map.  The red curve is a lognormal + power-law fit to the histogram above the range of column density affected by the uncertainty due to the choice of the map area.  The light gray area around N/N$_0$ = 1 shows the change in N-PDF following the change in the map area by $\pm$20\%, even though it is possible to completely sample the N-PDF by including the entire simulation cube.  Toward the high column density end, the colored power-law distributions in the power-law regime of the full N-PDF are the N-PDFs of independent ``leaf'' structures in the dendrogram.  The summation of N-PDFs of all ``leaf'' substructures is shown as the gray area in the power-law regime, making up most of the power-law component of the full N-PDF.}
\end{figure*}

Figure \ref{fig:cartoon_sim} shows that, at t = 0.6 t$_\text{free-fall}$, the 2.3pc by 2.3pc integrated column density map derived from the Enzo simulation has a lognormal component toward the low column density.  The shape of the N-PDF is consistent with the N-PDF of the full 4.6pc by 4.6pc column density map derived from the entire cube \citep{Burkhart_2015a,Collins_2012}.  Figure \ref{fig:cartoon_sim} also shows that, when we mimic the observational constraint on the map area by changing map size, the change in the N-PDF occurs at the low column density side of the peak column density \citep[consistent with observations presented in \S\ref{sec:obsv_ln} and Figure \ref{fig:cartoon_obsv} in this paper and in][]{Lombardi_2015}, even though in simulations a complete sampling is possible by including the entire cube.  There is no detection limit in simulations.  Fitting of the lognormal component is thus stable and can be shown to correlate with physical properties \citep*{Burkhart_2015a}.

\section{The Power-law Component and the Dendrogram Decomposition of the N-PDF}
\label{sec:pl}

\subsection{Observation}
\label{sec:obsv_pl}
\begin{figure*}[!b]
\centering
\includegraphics[scale=0.35]{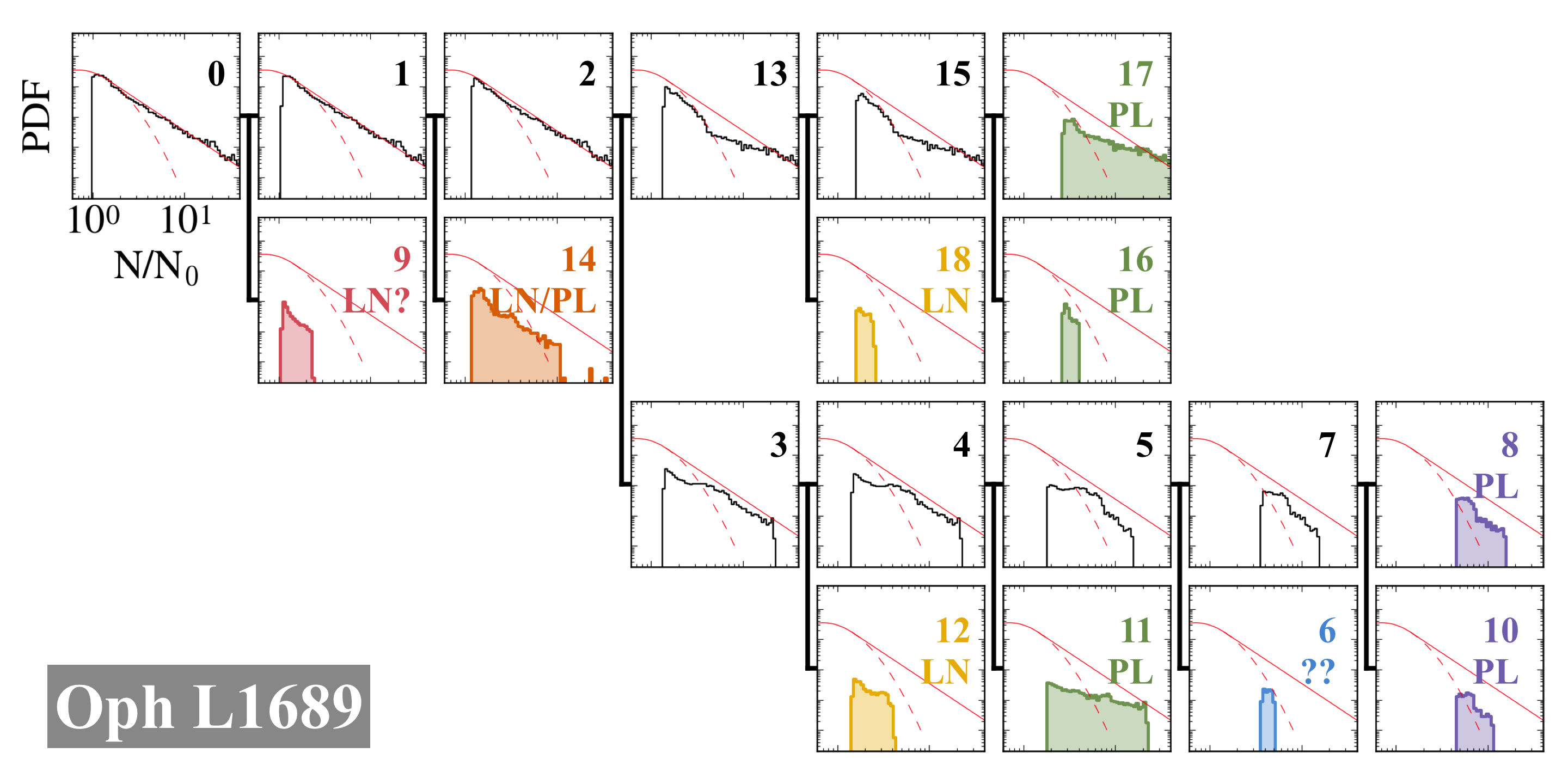}
\caption{\label{fig:fancy_obsv}The connection between dendrogram structures are shown as the thick lines connecting the panels, and the N-PDF of each structure is shown inside each panel.  The red curve in each panel is a lognormal + power-law fit to the N-PDF of the entire Oph L1689.  The ``leaf'' structures (\textit{i.e.} independent structures that cannot be broken down into smaller structures) in the dendrogram are color coded according to their levels in the dendrogram tree.  The numberings of the substructures are produced by \texttt{astrodendro}.  LN, LN/PL and PL denote the categories discussed in \S\ref{sec:sim_obsv} and stand for "lognormal," "lognormal + power-law" and "power-law."  See Figure \ref{fig:map_sim} for locations of the ``leaf'' structures on the integrated column density map.}
\end{figure*}

Figure \ref{fig:cartoon_obsv} shows that the N-PDF of Oph L1689 has a power-law component above N/N$_0$ $\sim$ 1.4 (the transition point).  In an attempt to understand the composition of the power-law component, we apply the dendrogram algorithm to find structures with physical sizes larger than 0.05 pc above a minimum value of N/N$_0$ = 0.95 \citep*[$A_K$ $\sim$ 0.8, consistent with the ``dense cloud'' definition suggested by][]{Lada_2010}, with \texttt{min\char`_delta} = 0.475 in the normalized unit.  (See \S\ref{sec:dendro} for the specifics of the \textit{dendrogram} algorithm and the setup parameters.)  The resulting Ophiuchus dendrogram is shown as the connections between panels in Figure \ref{fig:fancy_obsv}.

When we examine the N-PDFs of individual ``leaf'' structures (color coded according to their levels in the dendrogram in Figure \ref{fig:fancy_obsv}), we find that the N-PDFs of the individual ``leaf'' structures can be roughly categorized into three categories.  First, there are ``leaf'' structures such as Structure 9, 12, and 18 (marked with ``LN'' in Figure \ref{fig:fancy_obsv}), which sit at the lower levels in the dendrogram and have N-PDFs mostly in the lognormal regime of the entire cloud.  These are likely transient (unbound) over-densities, and their N-PDFs do not always look like a lognormal function because that the sample size is small.  Secondly, Structure 14 (marked with ``LN/PL'' in Figure \ref{fig:fancy_obsv}) could be at an early stage of gravitational collapse.  It sits at a low level in the dendrogram tree, and has an N-PDF consisting of both a lognormal component and a power-law component.  Lastly, there are several structures sitting toward the top of the dendrogram tree with N-PDFs almost entirely in the power-law regime of the entire Oph L1689.  Structure 6, 8, 10, 11, 16 and 17 belong to the last category (marked with ``PL'' in Figure \ref{fig:fancy_obsv}).  The N-PDF of each of these structures has a shape roughly resembling the power-law distribution.  Some of these structures (\textit{e.g.}, Structure 6, 8, and 10 in Figure \ref{fig:fancy_obsv}) span a smaller range of column density, while others (\textit{e.g.}, Structure 11 and 17 in Figure \ref{fig:fancy_obsv}) are much denser with narrow $^{13}$CO (1-0) line widths suggesting that they are gravitationally bound (see a full virial analysis of the dynamics to be presented in Chen et al.\ 2017, in preparation).  Notice that the slopes of the N-PDFs of the dense substructures are not necessarily the same as that of the entire Oph L1689.  The different slopes may indicate that the structures are undergoing gravitational collapse at different evolutionary stages \citep{Stutz_2015, Burkhart_2015b}.  The sum of all the leaf N-PDFs, given by the grey shaded area in Figure \ref{fig:cartoon_obsv} (and in Figure \ref{fig:cartoon} and Figure \ref{fig:cartoon_sim}), is a very good approximation to the observed PDF.

Following the nested structures of the dendrogram from the top-level ``leaf'' structures down to the ``branch'' structures, we find that the N-PDF of a ``branch'' structure containing several or more ``leaf'' structures with power-law N-PDFs has a power-law component similar to the power-law component of the N-PDF of the entire region.  (For example, see the N-PDF of the branch that contains Structure 12, 11, 6, 8, and 10 in Figure \ref{fig:fancy_obsv}.)  And, Figure \ref{fig:cartoon_obsv} shows that most of the pixels in the power-law component of the entire region are in the independent ``leaf'' structures in the dendrogram.  Since a power-law N-PDF is analytically expected from the self-similar gravitational collapse of a cloud \citep{Shu_1977}, these results suggest that the power-law component of an extended region is the summation of N-PDFs of dense, probably self-gravitating, substructures within the region.

\subsection{Simulation}
\label{sec:sim_pl}
\begin{figure*}[!b]
\centering
\includegraphics[scale=0.35]{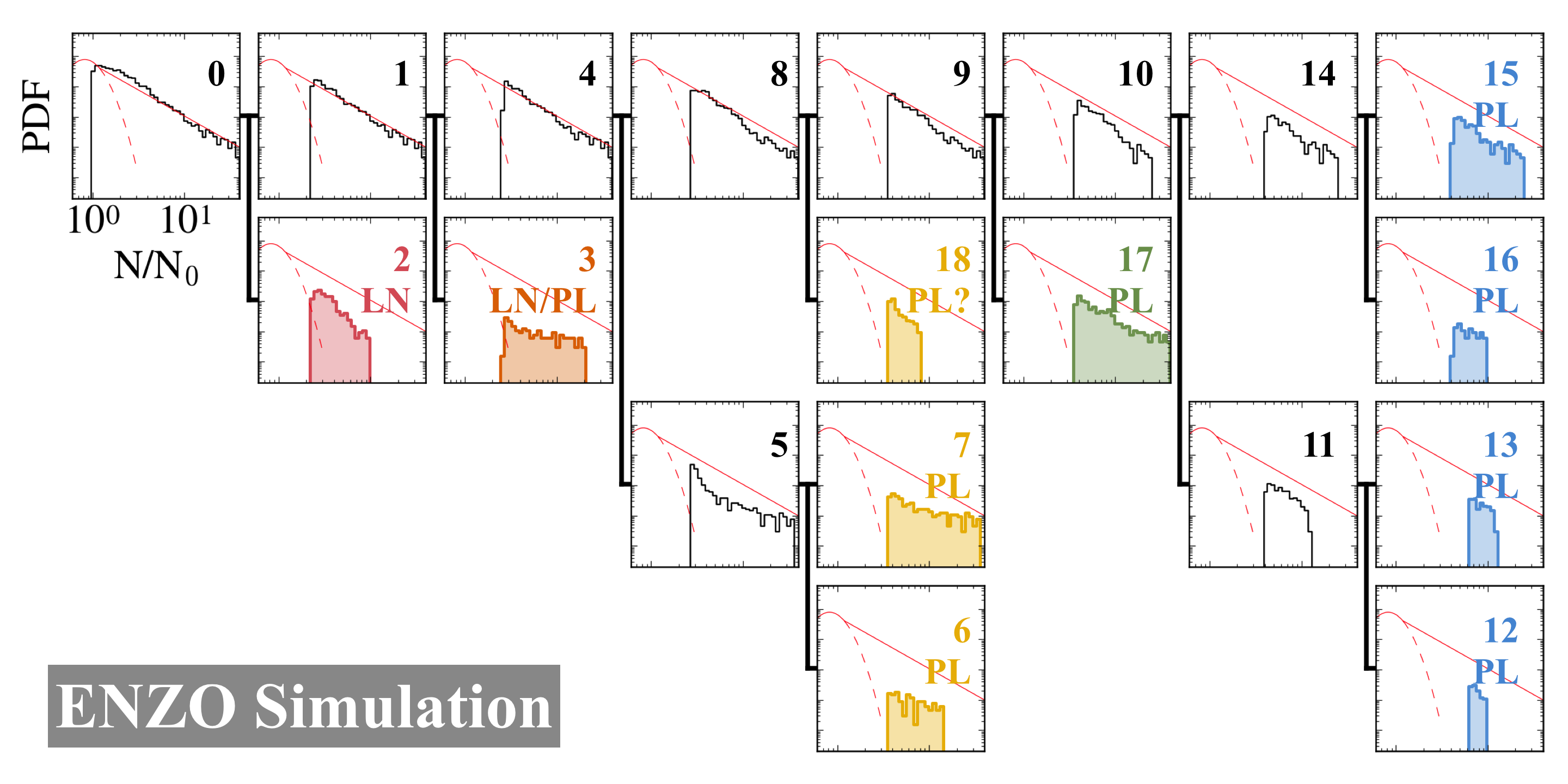}
\caption{\label{fig:fancy_sim} Similar to Figure \ref{fig:fancy_obsv}, but for the Enzo simulation.  The connections between dendrogram structures are shown as thick lines connecting the panels, and the N-PDF of each structure is shown inside each panel.  The red curve in each panel is a lognormal + power-law fit to the N-PDF of the entire 2.3pc by 2.3pc region.  The ``leaf'' structures in the dendrogram are color coded according to their levels in the dendrogram tree.  The numberings of the substructures are produced by \texttt{astrodendro}.  LN, LN/PL and PL denote the categories discussed in \S\ref{sec:sim_pl} and stand for "lognormal," "lognormal + power-law" and "power-law."  See Figure \ref{fig:map_sim} for locations of the ``leaf'' structures on the integrated column density map.}
\end{figure*}

Similar to observations (Figure \ref{fig:cartoon_obsv}), Figure \ref{fig:cartoon_sim} shows that the N-PDF of the Enzo simulation at t = 0.6 t$_\text{free-fall}$ has a power-law component above N/N$_0$ $\sim$ 1.2.  We then apply the dendrogram algorithm on the simulated column density map with the same set of setup parameters as that applied on observations.  (See \S\ref{sec:dendro} for details on the dendrogram setup parameters.)  The dendrogram is shown as connections between panels in Figure \ref{fig:fancy_sim}.  Notice that the dendrogram of the simulated column density map has a similar complexity as that of Oph L1689 (Figure \ref{fig:fancy_obsv} and Figure \ref{fig:fancy_sim}).

Figure \ref{fig:fancy_sim} shows the N-PDFs of substructures in the dendrogram.  The N-PDFs of all ``leaf'' structures seem to sit in the power-law regime of the lognormal + power-law model fitted to the N-PDF of the entire region (Equation \ref{eq:piecewise}; see \S\ref{sec:fitting} for details on the lognormal + power-law model).  However, we can still identify ``leaf'' structures that have N-PDFs with shapes of a lognormal distribution (Structure 2), a lognormal + power-law distribution (Structure 3), or a power-law distribution (Structure 6, 12, 13, 15, 16, 17 and 18).  The slopes of the power-law component of the denser substructures are different from each other and from the slope of the power-law component of the entire region.  And similar to observations, Figure \ref{fig:cartoon_sim} shows that most of the pixels in the power-law component of the entire region are in the independent ``leaf'' structures.  Most of these ``leaf'' structures also have power-law N-PDFs, albeit with different slopes.  And, similar to Oph L1689, the power-law component of the N-PDF of the selected region in the Enzo simulation is a summation of the N-PDFs of likely self-gravitating substructures within the cloud \citep{Shu_1977}.

\section{The Transition Point}
\label{sec:trans}
\citet*{Burkhart_2017} interpret the transition point as the density at which the turbulent energy density is equal to the thermal pressure and provide an analytic model of the transitional column density value \citep[see also][]{Myers_2015}.  In this section, we test this analytic model and investigate the physics behind the transitional column density value between lognormal and power-law distributions.  In the following paragraphs, we present the analytic model derived by \citet*{Burkhart_2017} assuming an equilibrium between the turbulent energy density and the thermal pressure at the transitional column density.  See \citet{Myers_2015} for a more general mathematical description of the transition point.

We consider a piecewise form of the N-PDF as in Equation \ref{eq:piecewise}. (See \S\ref{sec:fitting} for more details on the lognormal + power-law model.)  By assuming that the N-PDF at the transition point is continuous and differentiable, the transition point was derived in \citet*{Burkhart_2017} as:

\begin{equation}
\label{eq:continuity}
s_t=\frac{1}{2}\left(2\left|\alpha\right|-1\right)\sigma_s^2 ,
\end{equation}

\noindent where $s_t$ is the logarithmic normalized column density at the transition point, $\alpha$ is the slope of the power law tail, and $\sigma_s$ is the width of the lognormal component of the N-PDF.

In the strong collapse limit, where a well defined power-law tail is formed, the slope of the power-law component, $\left|\alpha\right|$, assumes a value of $\sim$ 1.5. In this limit, \citet*{Burkhart_2017} showed that the normalized column density at the transition point, N$_t$/N$_0$, can be expressed as a function of the sonic Mach number and the forcing parameter:

\begin{equation}
\label{eq:transition}
\frac{\text{N}_t}{\text{N}_0} \approx \left(1+b^2\mathcal{M}_s^2\right)^{A}\text{,}
\end{equation}

\noindent where $b$ is the forcing parameter, varying from $\approx$ 1/3 (purely solenoidal forcing) to 1 (purely compressive forcing), and $A$ = 0.11 is the scaling constant from volume density to column density \citep{Federrath_2008,Burkhart_2017}.  We can then test Equation \ref{eq:transition} by comparing this modeled transitional column density with the fitted transitional column density.  (See \S\ref{sec:fitting} for details on how we fit for the transition point.)  In the following subsections, we present the results of the tests in observations and simulations.

\subsection{Observation}
\label{sec:obsv_trans}
We compare the fitted transitional column density value to the modeled value \citep*[see Table \ref{table:trans} in this paper;][]{Burkhart_2017}.  Equation \ref{eq:transition} shows that the normalized transition column density, N$_t$/N$_0$, is dependent on the sonic Mach number, $\mathcal{M}$.  We fit Gaussian line profiles for spectra from FCRAO observations of $^{13}$CO (1-0) line emission \citep[the COMPLETE Survey,][]{Ridge_2006} and find an average Mach number of Oph L1689, $\mathcal{M}_s$ = 4.50$^{+1.43}_{-1.09}$ (Table \ref{table:trans}; see \S\ref{sec:obsv_turb} for details on the estimation of the sonic Mach number in observation).  Using Equation \ref{eq:transition}, we then derive the modeled transitional column density, (N/N$_0$)$_{\text{trans, model}}$ = 1.14$^{+0.12}_{-0.08}$ with the forcing parameter $b$ = 1/3 (purely solenoidal), and (N/N$_0$)$_{\text{trans, model}}$ = 1.40$^{+0.18}_{-0.15}$ with $b$ = 1 (purely compressive).  Since, in observations, we do not know the forcing parameter, $b$, which varies between 1/3 and 1, the modeled transitional column density, (N/N$_0$)$_{\text{trans, model}}$, has a large uncertainty and ranges between 1.14 and 1.40.

To obtain the fitted transitional column density from the N-PDF, we first derive a binned histogram representation of the N-PDF and fit the histogram to the lognormal + power-law model.  (See \S\ref{sec:fitting} for details on the $\chi^2$-fitting.)  For the L1689 data, we find a fitted transitional column density, (N/N$_0$)$_{\text{trans, fit}}$ = 1.43$\pm$0.14 (Table \ref{table:trans}).  Compared to the modeled values presented in the above paragraph, we can say, with some uncertainties, that the observed N-PDF has a transition point mildly more consistent with the analytic model with a purely compressive forcing ($b$ = 1).

\begin{table*}[ht]
\caption{\label{table:trans}Predictions of the column density at the transition point compared to the fitted values.}
\centering
\begin{tabular}{lccccc}
 & Mach Number  & Forcing Parameter\tablenotemark{a} & Modeled Transition & Fitted Transition\tablenotemark{b} & Whether the Prediction\\
 & $\mathcal{M}_s$ & $b$ & (N/N$_0$)$_{\text{trans, model}}$ & (N/N$_0$)$_{\text{trans, fit}}$ & Agrees with the Fit\tablenotemark{c}\\
\hline
\hline
Observation (Oph L1689) & \tablenotemark{d}4.50$^{+1.43}_{-1.09}$ & 1/3 & 1.14$^{+0.12}_{-0.08}$ & 1.43$\pm$0.14 & No\\
 & \tablenotemark{d}4.50$^{+1.43}_{-1.09}$ & 1 & 1.40$^{+0.18}_{-0.15}$ & 1.43$\pm$0.14 & Yes\\
\hline
Simulation (Enzo) & 8.14$^{+2.95}_{-1.81}$ & 1/3 & 1.26$^{+0.11}_{-0.08}$ & 1.18$\pm$0.12 & Yes\\
\end{tabular}
\begin{tablenotes}
\item[a]{\textbf{a.} The forcing parameter ranges from 1/3 (purely solenoidal forcing) to 1 (purely compressive forcing).}
\item[b]{\textbf{b.} The uncertainty is estimated based on the inherent uncertainties of Herschel observations.}
\item[c]{\textbf{c.} The two values are said to agree when the ranges enclosed by the uncertainties overlap.}
\item[d]{\textbf{d.} The Mach number is derived from the average velocity dispersion of Gaussian fits to the $^{13}$CO (1-0) molecular line emissions in the region.}
\end{tablenotes}
\end{table*}

\subsection{Simulation}
\label{sec:sim_trans}
\citet*{Burkhart_2017} have verified that the transitional column density can be described by the above analytic expression (\S\ref{sec:trans}), using a set of Enzo simulations with various setups of the sonic Mach number and the Alfv\'enic Mach number \citep*[see Figure 3 in][]{Burkhart_2017}.  Since it is possible to completely sample the lognormal + power-law distribution in simulations, \citet*{Burkhart_2017} followed Equation \ref{eq:continuity} \citep*[that is Equation 6 in][]{Burkhart_2017} and demonstrated that the fitted transitional column density matches the modeled value as a function of the fitted width of the lognormal component, $\sigma_s$, and the fitted slope of the power-law component, $\alpha$ (Equation \ref{eq:continuity}).

In this section, we present a separate verification of the analytic model of the transition point, using the 2.3pc by 2.3pc integrated column density map (Figure \ref{fig:map_sim}; see also \S\ref{sec:sim} for details on how the map was made).  Instead of modeling the transitional column density value based on the fitted width of the lognormal component and of the fitted slope of the power-law component \citep*[see Equation \ref{eq:continuity};][]{Burkhart_2017}, we follow Equation \ref{eq:transition} and model the transitional column density based on the Mach number, $\mathcal{M}_s$, and the forcing parameter of the simulations, $b$.  Since the Mach number and the forcing parameter are independent from the fitting of the N-PDF, the test presented below following Equation \ref{eq:transition} is stricter than (and \emph{independent from}) the one presented in \citet*{Burkhart_2017}, which follows Equation \ref{eq:continuity} and involves fitting the N-PDF on both sides of the equation.

We calculate the sonic Mach number, $\mathcal{M}_s$, based on the 3D velocity dispersion of the 2.3pc by 2.3pc by 4.6pc cube (a 2.3pc by 2.3pc map with a 4.6pc line of sight) from which the column density map was derived.  Since the gravity in the 2.3pc by 2.3pc region is particularly dominant (see Figure \ref{fig:map_sim} in which multiple high-density structures are identifiable), we expect a smaller sonic Mach number (less turbulent materials) than expected of the entire cube \citep[$\mathcal{M}_s$ = 9;][]{Collins_2012}.  For the 2.3pc by 2.3pc region shown in Figure \ref{fig:map_sim}, we find $\mathcal{M}_s$ = 8.14$^{+2.95}_{-1.81}$ (see \S\ref{sec:sim_turb} for details on the estimation of the sonic Mach number in simulation, and also \S\ref{sec:comparison} for a discussion on differences in the physical properties between Oph L1689 and the Enzo simulation used in this paper).  Knowing that the simulation has purely solenoidal forcing ($b$ $\approx$ 1/3), we then derive the transitional column density based on the analytic model (Equation \ref{eq:transition}), (N/N$_0$)$_{\text{trans, model}}$ = 1.26$^{+0.11}_{-0.08}$.  This is consistent with the fitted transitional column density, (N/N$_0$)$_{\text{trans, fit}}$ = 1.18$\pm$0.12 (Table \ref{table:trans}).  The result again verifies that the transition point of the observed N-PDF is well described by the analytic model proposed by \citet*{Burkhart_2017}, especially in simulations.

Table \ref{table:trans} gives an overview of comparisons between the fitted and modeled transitional column densities in observation and simulation.  We see that when the transition point between the lognormal and the power-law components of an N-PDF can be fitted, the analytic model could be potentially useful for deriving the dynamics of star-forming materials from the column density distribution.  Unfortunately, the ability to estimate the transitional column density in the lognormal + power-law model is limited by the uncertainty due to changing the map area used to derive the N-PDF in observations.  On top of the difficulty in fitting, the forcing parameter, $b$, is usually difficult to measure in observations, adding a large uncertainty to the modeled transitional column density \citep[see recent attempts by][]{Orkisz_2017,Herron_2017,Otto_2017}.


\section{Discussion}
\label{sec:discussion}

Our study highlights the importance of comparing observations and simulations.  For example, the lognormal portion of the PDF in observations suffers biases, such as boundary effects and unresolved foreground/background contributions \citep{Schneider_2015a,Lombardi_2015}. Simulations provide an avenue to study these effects as the density in the simulations is completely sampled due to mass conservation and periodic boundary conditions. However, the simulations are missing important physical effects such as feedback from stars \citep{Offner_2015}, non-isothermal effects \citep{Nolan_2015} and non-ideal MHD effects \citep{Meyer_2014,Burkhart_2015c}.  No single simulation is able to capture all the physical processes and scales involved in star formation.  Simulations, therefore, can serve only as a general reference to interpret the observations.  

One aspect not addressed in our study of the anatomy of the N-PDF is the influence of the magnetic field.  The effect of the magnetic field on the shape and behavior of the PDF of gravitating turbulent clouds has been studied in the past \citep{Burkhart_2012,Mocz_2017,Kritsuk_2011,Collins_2012,Federrath_2012}.  In general, the lognormal portion of the N-PDF is not strongly affected by changing the magnetic field strength \citep{Burkhart_2009, Burkhart_2012,Collins_2012,Federrath_2012,Burkhart_2015a}.  However, studies have shown that the strength of the magnetic field can alter the slope of the power-law tail portion of the N-PDF \citep{Burkhart_2015a,Mocz_2017}.  These studies found that a higher magnetic field strength produces an N-PDF with a steeper power-law tail slope since the magnetic field inhibits the collapse.


\subsection{The dendrogram analysis and the N-PDF of an independent structure}
\label{sec:discussion_structure}
The analysis presented in this paper shows that the power-law component of the N-PDF is the sum of individual substructures with power-law PDFs.  Past a transition point where the shape of the PDF changes from lognormal to power-law, individual dendrogram structures show clear power-law forms, while at or below the transition point the column density PDFs of substructures primarily take on a lognormal form (e.g. structures 11, 14, 17 in Figure \ref{fig:fancy_obsv}).

One notable deviation from the above picture is Structure 6 (see Figure \ref{fig:map_obsv} and Figure \ref{fig:fancy_obsv}) in the dendrogram analysis of the N-PDF of Oph L1689.  Structure 6 sits above the transition point, so we would expect that its N-PDF should take on a power-law shape.  However, the N-PDF of Structure 6 is visually indistinguishable from a lognormal distribution.  Despite the fact that there are only $\sim$ 2$\times$10$^2$ pixels in Structure 6 and that the sampling of the N-PDF is thus probably incomplete, we want to ask the question: Does the lognormal shape of the N-PDF of Structure 6 mean that the substructure is, in fact, not gravitationally bound?  Ongoing star formation is evident in this region within a time scale $\sim$ 5$\times10^5$ years, since there are multiple young stellar objects nearby \citep{Gutermuth_2009}.  Feedback from active star formation could have shredded the star forming materials in the region into smaller, unbound pieces like Structure 6.  The shredded pieces would have N-PDFs that become increasingly steep, potentially mimicking a power-law distribution at high column density.  (Similar effects of feedback on the N-PDF shape have been observed at cloud scales by \citet{Schneider_2015a}.)  We also note that the Enzo simulation we compare to (Figure \ref{fig:fancy_sim}) lacks a similar lognormal substructure at these high densities, and that there is no feedback in the Enzo simulation.  Local environmental effects such as feedback therefore most likely play a role in the unusual shape of structure 6.


If an independent structure (a leaf structure in the dendrogram) is axially symmetric (\textit{e.g.}, a filamentary structure) or radially symmetric (\textit{e.g.}, a spherical structure), \citet{Myers_2015} pointed out that the N-PDF is the inverse function of the column density profile along a cut perpendicular to the axis of symmetry or in the radial direction, respectively.  \citet{Myers_2015} observed that, for a filamentary structure with a Plummer-like profile perpendicular to the axis of the filament \citep[typical of star forming filamentary structures;][]{Arzoumanian_2011}, the inverse function of the Plummer-like column density profile deviates from the lognormal+power-law model of the N-PDF.  \citet{Myers_2015} suggested that this is likely due to the nearly constant column density within the first thermal scale length.  Similarly, the inverse function of the narrow N-PDF of structure 6 in Figure \ref{fig:fancy_obsv} would have a flat shape, consistent with a nearly constant column density profile.  Is the thermal length scale then the scale limit for the N-PDF to be representative of the gravitational effect?  A more detailed analysis of the velocity structure of dendrogram features such as Structure 6 in Oph L1689 (see Figure \ref{fig:fancy_obsv}) is needed to answer this question.


\section{Conclusion}
\label{sec:conclusion}
We present the anatomy of the column density PDF (N-PDF) of star forming molecular clouds in both observations and simulations.  By assuming a lognormal + power-law model, we examine how the lognormal component, the power-law component, and the transition point could be useful for estimating the dynamical properties of a star forming region.  We also examine the uncertainties that could affect the N-PDF analysis.

With the help of the dendrogram algorithm, \emph{we demonstrate that the power-law component of an N-PDF is primarily a summation of N-PDFs of substructures inside the star forming cloud}.  Most of these substructures show N-PDFs following power-law distributions, with power-law indices different from the N-PDF of the entire region.  The power-law shapes and varying indices of N-PDFs suggest that these substructures could be going through different stages of gravitational collapse.

The analytic prediction of a transition point between lognormal and power-law components proposed by \citet*{Burkhart_2017} is verified independently (within uncertainties) in this work using data from both observations and simulations.   We show that the transition point is generally well described by the analytic model \citep*[see Equation \ref{eq:transition} in this paper;][]{Burkhart_2017}.  The result suggests that finding the transitional column density value in a N-PDF could potentially provide information on the dynamics of the cloud.  In particular, Equation \ref{eq:transition} could potentially be useful for getting a general estimate of the sonic Mach number, even though it is usually difficult to determine the transition point and the forcing parameter in observations \citep[see a recent attempt by][]{Orkisz_2017}.



Based on the results presented in this paper, we give the following suggestions for future studies involving the N-PDF analysis.  First, we do not suggest analyzing the dynamics based solely on fits to the lognormal component of the N-PDF derived from observations of dust tracers.  Second, measuring the column density at the transition point between the lognormal and the power-law components could potentially provide information on the dynamical properties of a star forming region, but only when the transitional column density and the forcing parameter can be estimated.  \emph{Lastly but most importantly, we recognize that the power-law component of an N-PDF is a summation of N-PDFs of substructures, which are likely going through various stages of gravitational collapse.  Based on the results presented in this paper, we suggest combining the N-PDF analysis with the dendrogram algorithm to obtain a more complete picture of the effects of global and local environments, including gravity, turbulence, the magnetic field, and the feedback from star formation, in future attempts to analyze the column density structures of a star forming region.}

\acknowledgments
B.B. acknowledges support from the NASA Einstein Postdoctoral Fellowship. 
Computer time was provided through NSF TRAC allocations TG-AST090110 and TG-MCA07S014. The computations were performed on Nautilus and Kraken at the National Institute for Computational Sciences (\url{http://www.nics.tennessee.edu/}).

\bibliography{main.bib}

\end{document}